\documentclass[french]{article-hermes}
\usepackage{graphicx}

\begin{document}



\title[Vers un déploiement transactionnel]
      {Gestion transactionnelle \\de la reprise sur erreurs dans le déploiement}

\author{Cristina Marin \andauthor{Noureddine Belkhatir} \andauthor{Didier Donsez}}

\address{%
LSR - Adèle, 220 Rue de la Chimie\\
Domaine Universitaire, BP 53\\
38041 Grenoble France\\[3pt]
\{cristina.marin,noureddine.belkhatir,didier.donsez\}@imag.fr}

\resume{Avec le développement des réseaux et de l'Internet, la problématique du déploiement automatisé à large échelle est devenue de plus en plus cruciale. Le déploiement est un procédé logiciel complexe couvrant plusieurs activités allant de la configuration jusqu'à la désinstallation du logiciel. Pendant l'exécution d'un procédé de déploiement, des exceptions peuvent être rencontrées qui mettent le site dans un état incohérent. Pour les résoudre, nous proposons une approche à base des concepts transactionnels pour décrire les actions à entreprendre lorsqu'une situation exceptionnelle est rencontrée à l'exécution. L'approche proposée aide à respecter la cohérence du site en conservant une partie du travail déjà effectué par le procédé. Cet article présente notre approche et une expérimentation faite dans un système de déploiement issu de la recherche académique.}

\abstract{With the development of the networks and the Internet, the problems of automated deployment on broad scale became increasingly crucial. Software deployment is a complex process covering several activities going from the configuration to the retirement of a software product. During the execution of a deployment process, exceptions can be met which put the site in an incoherent state. To solve them, we propose an approach based on transactional concepts which describes the actions to be undertaken when an exceptional situation is met during the deployment process. The approach guaranties the respect of the site's consistency by preserving part of the work already carried out by the process. This article presents our approach and an experimentation made in an academic deployment system.}

\motscles{Déploiement automatisé, procédé de déploiement, exceptions, transactions, cohérence}

\proceedings{DECOR'04, Déploiement et (Re)Configuration de Logiciels}{199}

\maketitlepage

\section{Introduction}

Le déploiement est devenu dans ces dernières années une activité d'une importance croissante pour les entreprises qui utilisent des logiciels complexes et de taille importante. Le déploiement à large échelle (des milliers des sites) de ce type d'application pose des problèmes qu'il est difficile de résoudre sans l'aide d'un support automatisé. 

Le domaine du déploiement logiciel s'intéresse au développement et à la conception de systèmes destinés à fournir un support automatisé au déploiement massif de logiciels. Il essaie de définir des solutions fiables et performantes et des mécanismes  automatisés pour traiter le déploiement. En conséquence, toutes les activités du déploiement \cite{Car98} peuvent être exécutées avec le minimum d'intervention de la part de l'utilisateur. 

Il existe de nombreuses approches et de nombreux outils de déploiement, mais très peu permettent de couvrir entièrement le cycle de vie du déploiement en imposant des contraintes fortes, comme la plate-forme d'exécution ou le type de l'application à déployer. Ils sont aussi construits de manière ad-hoc. La solution proposée dans notre approche consiste à introduire des abstractions. Une des abstractions est le modèle décrivant le procédé de déploiement. Ce modèle est exécutable. Le système de déploiement sera ainsi capable de coordonner l'exécution des activités de déploiement. 

Cependant un procédé de déploiement peut rencontrer à l'exécution des situations exceptionnelles qui peuvent mettre en échec sa terminaison et la cohérence des sites sur lesquels les logiciels sont déployés. La cohérence des sites est une caractéristique importante du déploiement. Nous l'exprimons par deux propriétés : la réussite et la sûreté\cite{Par00}. Ces deux propriétés du déploiement expriment qu'après l'exécution d'un procédé de déploiement l'application qui vient d'être déployée fonctionne comme spécifié et que le déploiement de la nouvelle application n'endommage pas le fonctionnement des autres applications déjà installées sur le site. 

Les outils de déploiement actuels traitent de manière identique les exceptions apparues à l'exécution en remettant le site dans l'état antérieur au déploiement. Notre approche consiste à définir un modèle de déploiement réactif qui analyse l'état du site et du procédé de déploiement à l'apparition d'une situation exceptionnelle en cours d'exécution et qui choisit une réaction appropriée pour préserver la cohérence du site et une partie du travail déjà effectué. 

L'article est organisé de la manière suivante. Nous présentons d'abord les concepts principaux du déploiement en réalisant ensuite une synthèse des outils de déploiement existants. La partie état de l'art présente la technologie des procédés exécutables et les différents modèles de transactions existants. Dans la section \ref{sect1}, nous détaillons notre problématique et présentons notre approche. Afin de le valider, nous présentons une expérimentation faite dans un système de déploiement académique. Enfin, nous concluons et présentons les perspectives de ce travail.

\section{État de l'art}
Cette section présente un survol des trois principaux domaines de notre travail. Il s'agit du thème central de notre travail de recherche - le déploiement  et aussi de la technologie des procédés exécutables ainsi que des transactions. Pour chacun d'eux, nous présentons brièvement les principaux concepts et les travaux le concernant. 

\subsection{Déploiement}
Le déploiement logiciel fait référence à toutes les activités et leurs interactions pendant le cycle de vie d'une application\cite{Car98}. Ces activités concernent le versionnement, l'installation, la désinstallation, la mise à jour, l'adaptation(la réconfiguration), l'activation et la désactivation d'une application.

Dés qu'une nouvelle version d'un composant logiciel est créée chez un producteur (\textsl{versionnement}), elle peut être \textsl{installée} sur une ou plusieurs sites du client. Pour pouvoir l'exécuter, l'application doit être \textsl{activée}. Quand le producteur dispose d'une nouvelle version d'un composant de l'application, celui-ci peut être \textsl{mis à jour}. Si un changement est produit dans l'environnement d'exécution de l'application (par exemple le changement de la carte vidéo utilisée par l'application), l'application peut être \textsl{réconfigurée}. Pour pouvoir effectuer des mises à jour ou des adaptations, l'application peut être \textsl{désactivée}. Enfin, quand un utilisateur n'a plus besoin d'une application installée, celle-ci peut être \textsl{désinstallée}. 
Cette présentation du cycle de vie du déploiement montre qu'il est composé de plusieurs activités complexes qui peuvent avoir éventuellement des liaisons entre elles. Chaque activité exécute une tâche précise et difficile. 

Les technologies industrielles actuelles liées au domaine du déploiement logiciel peuvent être classifiées en plusieurs catégories\cite{Car98}: \textsl{installateurs} (InstallShield X - ou Java Web Start), \textsl{gestionnaires de} \textit{packages} (RPM de Linux RedHat, 
SD de HP-UX), \textsl{systèmes de gestion des applications} (TME-10 de Tivoli), \textsl{standards de description de systèmes}, \textsl{systèmes de transfert de contenu}. Une étude plus détaillée sur ces catégories d'outils peut être trouvée dans \cite{Mar04}.

Ces produits ont été classifiés selon les problèmes du déploiement qu'ils traitent.  Ainsi, même si les solutions proposées sont efficaces, elles restent des solutions partielles qui ne couvrent pas tout le cycle de vie du déploiement. Parmi ces produits, certains couvrent une grande partie du processus de déploiement, mais imposent des contraintes sur le type de l'application à déployer (Java Web Start déploie seulement des applications web Java) ou de la plate-forme d'exécution (InstallShield accepte que des applications Windows). Un autre désavantage pour ces outils est qu'en cas d'erreurs à l'exécution ils défont tout le travail déjà effectué. Ainsi, le site est mis dans l'état antérieur à l'exécution du procédé de déploiement. Nous considérons cette approche pas toujours convenable. Un exposé des motifs qui nous ont conduit vers cette approche est détaillé dans la section \ref{sect1}.

D'autres solutions plus académiques proposent de définir des environnements de déploiement qui couvrent entièrement le cycle de vie du déploiement. Parmi ces approches les plus intéressantes sont SoftwareDock\cite{Hall}, qui est le résultat d'un travail de recherche de l'Université du Colorado, et \cite{ORYA3}, environnement de déploiement générique développé dans l'équipe Adèle du laboratoire LSR de Grenoble. Ces approches ont été matérialisées par le développement d'environnements de déploiement capables de gérer l'exécution des procédés de déploiement dans une architecture distribuée à grande échelle (SD) ou bien des environnements adaptables aux diverses politiques de déploiement qu'une entreprise peut définir (ORYA). Les technologies utilisées pour l'automatisation des activités sont différentes : des systèmes agents mobiles(SD) pour coordonner les activités du déploiement ou une fédération d'outils existants (ORYA) et un moteur de procédés exécutables (APEL). 

\subsection{Transactions}
Les transactions ont été introduites dans les systèmes de gestion de données pour fiabiliser les opérations pour des applications bancaires et les applications de réservation de billets d'avion (SABRE) et de train (SOCRATES). Dans ces systèmes, les applications se caractérisent généralement par une courte durée d'exécution (de l'ordre de la dizaine de seconde). La transaction est considérée comme un enchaînement ordonnancé des opérations simples (\textit{read/write}) respectant quatre propriétés, connues sous le nom de propriétés ACID.

Le concept de transaction a ensuite intéressé les concepteurs d'applications de conception assistée par ordinateur ou bien d'applications de gestion des documents. Ces applications se caractérisent quand à elles par la nécessité d'une durée longue d'exécution pour une transaction et par le besoin de coopération entre transactions. Les propriétés ACID se montrent alors trop restrictif.

Pour dépasser cet obstacle, plusieurs modèles dits avancés de transactions ont été proposés\cite{Elg}. L'idée est de relâcher les propriétés de l'atomicité et de l'isolation. Pour cela, une transaction longue est décomposée en plusieurs sous-transactions. La division va jusqu'à ce que les sous-transactions obtenues soient des opérations indivisibles (simples). C'est le cas des transactions coopératives, ou le cas de Sagas où les sous-transactions sont indépendantes les unes des autres.

Pour relâcher l'isolation, ces modèles permettent aux sous-transactions de rendre visible le résultat des sous-transactions, comme c'est le cas avec les transactions emboîtées ouvertes (ONT) ou les transactions Sagas. Pour récupérer un état cohérent des données en cas d'erreur, des mécanismes de contingences et des compensations sont utilisés(transactions emboîtées, des alternatives du modèle Sagas).

Le modèle ConTract s'intéresse aux longues durées d'exécution. Il propose de contrôler l'exécution des activités de longues durées avec une définition des étapes à exécuter pour garantir la réussite de l'activité.  

\subsection{Procédés}
La technologie des procédés a pour but de supporter l'exécution des procédés en fournissant les moyens de modéliser, d'analyser, d'améliorer, de mesurer et d'automatiser les activités de production.

La notion de \textit{modèle de procédé} a été définie comme une représentation abstraite d'une famille de procédés. Un modèle d'un procédé décrit un procédé en termes d'entités qui sont les activités et leur enchaînement, les ressources, les agents qui exécutent des rôles. Chacune de ces entités peut être elle-même modélisée. Pour décrire les procédés, il est nécessaire de disposer d'un \textit{langage de modélisation} (PML)\footnote{Process Modelling Language}. Tous les procédés exécutables d'une famille nécessitent un environnement qui coordonne leur exécution (PSEE)\footnote{Process Sensitive Software Engineering Environment}. Un tel environnement est contrôlé par un moteur de procédés (\textit{Process Engine}). Celui-çi a pour but de contrôler les flots de données qui transitent durant le procédé entre les différents acteurs intervenant à l'exécution du procédé. 

Dans ces dernières années, les systèmes de gestion des procédés sont utilisés dans une multitude d'applications. Ainsi peut être expliquée l'apparition d'une gamme assez variée de produits commerciaux: ProcessIT (AT\&T), InConcert (Xerox) ou FlowMark (IBM) sont seulement quelques exemples. Le problème est que tous ces produits ne résoudrent pas le traitement des exceptions. Pour cela, les chercheurs ont proposé d'utiliser les transactions pour modéliser un procédé transactionnel\cite{Sheth}.

Pour définir un modèle de procédé transactionnel, plusieurs approches ont été proposées basées sur une vision différente des projections des concepts. Une première approche est celui qui essaie d'ajouter à un modèle de procédé les concepts transactionnels pour assurer la fiabilité, la consistance de l'exécution. Dans l'autre approche, le modèle de transaction avancée est enrichi des concepts des procédés pour augmenter la fonctionnalité de l'application en terme de sûreté de l'exécution et des données. Les techniques sont utilisées de manière différente. Par exemple, la technique de reprise en arrière basée sur la compensation a deux modalités d'utilisation. D'une part, l'approche statique qui consiste de définir le modèle de compensation dans la phase de définition du procédé, soit comme partie integrante du procédé\cite{Alonso}, soit comme un modèle à part\cite{Muhlberger}. L'approche dynamique suppose la génération du plan compensatoire à l'exécution, mais son applicabilité reste trop complexe.

\section{Déploiement transactionnel}\label{sect1}
Un système de déploiement fiable et performant devra couvrir le cycle de vie du déploiement tout en tenant compte aussi des problèmes posés par les domaines adjacents du déploiement (les réseaux, la sécurité). La coordination et l'exécution des activités du déploiement peuvent être supportées par la technologie des procédés exécutables. 

Lors de l'exécution d'un procédé de déploiement des situations exceptionnelles peuvent être rencontrées. Celles-ci mettent le succès du procédé en péril et peuvent endommager la cohérence du site sur lequel le logiciel est déployé. La cohérence est exprimée par l'intermède de deux propriétés introduites par Parish dans \cite{Par00} : la \textbf{réussite} et la \textbf{sûreté}. Un procédé de déploiement est considéré comme réussi si l'application déployée fonctionne après l'installation ayant tous ces composants installés. La sûreté d'un procédé de déploiement est garantie si toutes les applications déjà installées sur un site continuent de fonctionner de la même manière après l'exécution du procédé. Pour pouvoir la garantir, un système de déploiement doit contenir des modèles pour exprimer les réactions à exécuter lorsqu'une exception survient en cours d'exécution.

Notre travail propose un modèle de traitement des situations exceptionnelles pour garantir les deux propriétés mentionnées. 

\subsection{Problématique}\label{A}
Les situations exceptionnelles rencontrées à l'exécution d'un procédé de déploiement peuvent être assez hétérogènes. Elles peuvent être causées par des pannes réseaux, par exemple l'échec du transfert d'un paquetage contenant un fragment d'une application vers un des sites sur lesquels il doit être déployé, des problèmes d'incompatibilité entre les composants installés sur un site ou des problèmes d'incompatibilité matérielle d'une application avec la plate-forme sur laquelle elle sera déployée. 

Le domaine du déploiement logiciel traite le cycle de vie d'une application. Une application est composée de plusieurs paquetages qui ont des dépendances entre eux à l'installation et/ou à l'exécution. Les composants d'une application peuvent avoir des différents degrés d'importance pour le fonctionnement d'une application. Prenons par exemple, le cas d'un logiciel très connu: Microsoft Word. Un composant obligatoire à l'exécution est l'éditeur de texte, toutefois le composant Dictionnaire n'est pas obligatoire d'être installé pour que l'application fonctionne. Ainsi, si à l'exécution d'un procédé de déploiement d'une application à composants, les composants obligatoires ont été installés lorsqu'une exception est levée, le procédé est mis en échec. Dans ce cas, plusieurs actions seront à envisager. Premièrement, la stratégie adoptée par tous les outils existants peut être appliquée; de cette manière, les composants installés devront être désinstallés et le site retrouverait un état cohérent. D'autre part, pour gagner en performance en terme de temps, comme tous les composants obligatoires de l'application ont été installés, le procédé de déploiement pourra être considéré comme réussi. Le déploiement des composants optionnels (tel que le dictionnaire) peut être considéré comme opération non-critique pour un procédé de déploiement, fait qui donnera un état de cohérence partielle du site.

De nos jours, un système de déploiement a une architecture distribuée qui contient plusieurs serveurs d'applications offrant les paquetages des applications à déployer, plusieurs sites sur lesquels une application devra être déployée et un serveur de déploiement qui gère l'exécution. Notre procédé de déploiement est composé de trois activités: premièrement un serveur qui contient le paquetage de l'application est trouvé, puis le paquetage est transféré vers le site client où il est enfin installé. Si pendant l'exécution du transfert (activité 2) le serveur d'applications tombe en panne, le déploiement est mis en échec. Ainsi, la réussite ne pourra pas être assurée. Comme souvent un paquetage n'est pas disponible seulement sur un seul serveur d'applications, on peut tout de même garantir la réussite du déploiement en suivant une autre voie d'exécution. Le procédé ne doit pas être abandonné, mais au contraire repris en complétant le chargement depuis un serveur d'applications miroir ou de secours.

Un autre exemple significatif est le cas d'un procédé de déploiement multiple. Prenons le cas d'un composant logiciel qui doit être installé simultanément sur un millier de sites. Si l'exécution du procédé n'est réussie que sur une partie importante des sites, le système de déploiement devra prendre une décision pour traiter cette situation exceptionnelle. Il pourrait appliquer la stratégie \textsl{tout ou rien}, mais dans ce cas la perte de performance sera trop significative. Il pourrait d'une autre coté ignorer l'échec du déploiement sur les sites manquantes et choisir ainsi de revenir plus tard et de réessayer d'installer le composant sur ces sites. Ainsi, les sites sur lesquels le déploiement est réussi pourront bénéficier du paquetage installé.

Les exemples que nous avons énoncés montrent le besoin d'avoir des mécanismes de reprise sur panne lorsqu'une exception apparaît en cours d'exécution d'un procédé de déploiement. Ces mécanismes doivent prendre en compte les critères de performance du système ainsi que celui de la cohérence du site. Dans les sections suivantes, nous décrivons le modèle que nous proposons pour traiter les problèmes posés par les erreurs sur un procédé de déploiement.

\subsection{Les propriétés ACID pour un procédé de déploiement}
Pour intégrer le concept de transaction dans le domaine du déploiement, nous avons étudié l'applicabilité des modèles de transactions existants sur un procédé de déploiement. Pour cela, il était important de caractériser l'exécution d'un procédé de déploiement. Le modèle d'exécution nécessaire pour notre procédé de déploiement possède les caractéristiques suivantes: il peut avoir une longue durée d'exécution (des heures,  même des jours); il est composé de plusieurs activités qui peuvent être exécutées en séquentiel (une installation d'une application sur un seul site) ou en parallèle (un procédé de déploiement multiple, dans lequel une application est déployée simultanément sur plusieurs sites). Ces activités peuvent être complexes en terme de ressources utilisées (volume de données transférées,CPU). 

Par rapport au déploiement, les propriétés ACID d'un procédé de déploiement vu comme une transaction pourront être définies de la manière suivante. La propriété d'atomicité demande l'exécution correcte de toutes les opérations composant le procédé pour que celui-ci puisse valider son exécution. Les contraintes demandées d'être respectées en garantissant la propriété de la cohérence sont celles déjà mentionnées - la réussite et la sûreté. Une exécution en respectant la propriété de l'isolation demande qu'aucune modification intermédiaire ne soit pas rendue visible pour les autres procédés s'exécutant en parallèle. Pour respecter la durabilité, toutes les modifications validées par des procédés exécutés avec succès devront persister dans le système.

Une exécution ACID n'est pas appropriée pour un procédé de déploiement. En cas d'erreur, les propriétés ACID imposent de défaire. Un autre procédé s'exécutant en même temps ou même des applications installées sur le site devront attendre la fin du procédé pour pouvoir utiliser des données (des fichiers, des composants logiciels) verrouillées par celui-ci. Les propriétés qui devront être assurées à l'exécution sont la cohérence (exprimée par les deux propriétés mentionnées) et la durabilité. L'isolation et l'atomicité ne sont pas nécessaires pour le procédé de déploiement mais peuvent être demandées pour une activité spécifique du procédé. 

\subsection{Procédé transactionnel de déploiement}
Nous proposons une approche basée sur les concepts issus des modèles de transactions avancées pour traiter les problèmes qui apparaissent au cours d'exécution d'un procédé de déploiement. L'exécution du procédé respectera ainsi les propriétés CD. Notre approche propose des solutions pour traiter les éventuelles erreurs. Comme un procédé de déploiement est un enchaînement de plusieurs activités, nous introduisons nos concepts au niveau activité. Dans ce qui suit, nous présentons les concepts introduits pour exprimer le comportement transactionnel sur un procédé de déploiement.

\textbf{Activité critique/non-critique.} Les activités composant un procédé de déploiement peuvent avoir différents degrés d'importance. Ainsi, le procédé peut contenir aussi bien des activités obligatoires à être exécutées (c.à.d activités critiques) que des activités non-critiques. Nous définissons une \textsl{activité non-critique} comme étant une activité qui ne doit pas être obligatoirement terminée avec succès pour que le procédé de déploiement soit réussi. 

\textbf{Point de reprise.} Une des propriétés qui doivent être assurées par un procédé de déploiement est la cohérence du site. Un procédé de déploiement en exécution peut avoir des étapes intermédiaires dans lesquels la cohérence du site est respectée aussi que des étapes dans lesquels elle ne l'est pas. Pour caractériser les étapes cohérentes d'un procédé de déploiement, nous introduisons la notion de point de reprise.  Un point de reprise est associé à une activité à la fin de laquelle l'état du site est cohérent. Il est utilisé comme point de retour en arrière en cas d'erreur à l'exécution pour que les modifications d'un procédé en échec ne soient pas défaites en totalité.

\textbf{Activité de contingence.} Une exception apparue pendant l'exécution d'un procédé de déploiement peut mettre en échec sa réussite. Toutefois, il existe des situations pour lesquelles, un autre chemin que celui de l'activité échouée peut aboutir avec succès.
Cette nouvelle façon de réaliser la tâche d'une activité en échec réalise la tâche d'une autre activité alternative, que nous appelons \textsl{activité de contingence}. 

\textbf{Activité de compensation.} Pour défaire en cas d'exception les modifications des activités terminées avec succès, nous avons défini des activités de compensations. Une activité de compensation défait les modifications validées de l'activité à laquelle elle est associée.

Tous ces concepts que nous avons défini pour un procédé de déploiement lui confèrent un comportement transactionnel. Ainsi, lorsqu'une erreur apparaît, l'échec est ignoré par le procédé si l'activité échouée est une activité non-critique. Pour traiter cet échec et afin d'assurer la réussite, une activité de contingence est exécutée. Puis, le procédé peut être poursuivi normalement. Si l'activité de contingence échoue, afin de garder une partie du travail déjà effectué et de retrouver un état cohérent du site pour garantir la sûreté, un point de reprise du procédé est utilisé. Pour retourner le procédé en arrière vers ce point, les compensations sont exécutées. Ainsi, les techniques de reprise sur panne transactionnelles utilisées sont celles des récupérations en avant et en arrière. 

De cette manière, un système utilisant ces concepts pour un procédé de déploiement est plus efficace grâce à l'introduction des notions d'activité non-critique et activité de contingence aidant de ne pas abandonner le procédé en présence d'une erreur. 

\section{Meta-modèle de procédé transactionnel}
\begin{figure*}[ht]
 \setlength{\fboxsep}{2mm}
 \fbox{
  \parbox{116mm}{
   \vspace{-20mm}
   \begin{center}
    \small
    \includegraphics[width=0.7\textwidth,angle=-90]{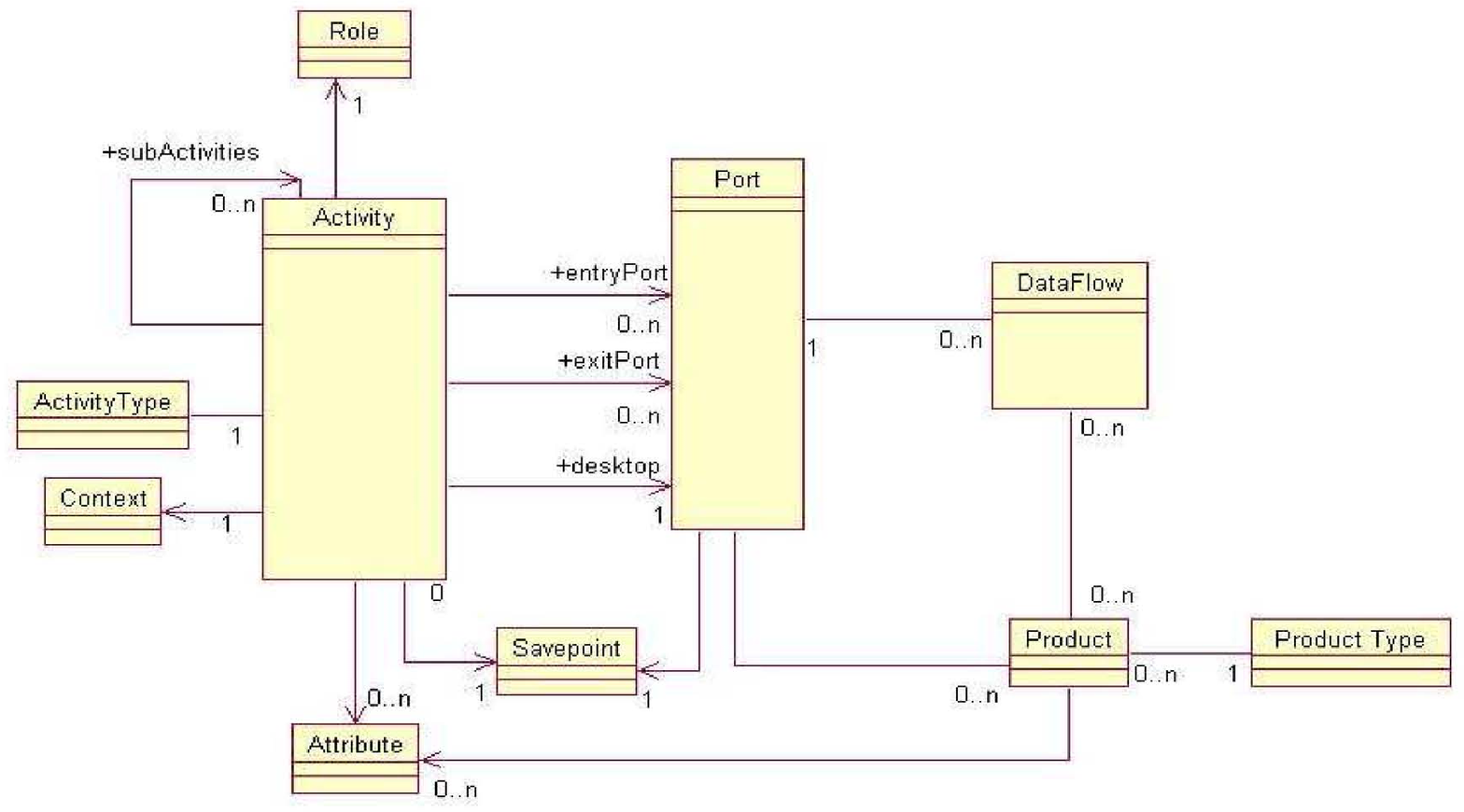}
   \end{center}
   \vspace{-20mm}
  }
 }
 \caption{Le meta modèle de procédé transactionnel}\label{fig:1}
\end{figure*}

La figure \ref{fig:1} représente le méta modèle intégrant des concepts transactionnels dans un modèle de procédés. Le modèle de procédé schématisé contient une description des entités composant un procédé de déploiement. Un procédé consiste d'un ensemble d'activités, simples ou composites(constituée d'un enchainement de plusieurs activités). Chaque activité exécute un rôle (\textit{Role}) et a des attributs (\textit{Attribute})(par exemple, l'état à l'exécution). Pour communiquer avec les autres activités du procédé, les activités possèdent des ports (\textit{Port}) qui reçoivent les produits (\textit{Product}). Les produits représentent les informations nécessaires pour que l'activité accomplie sa tâche, par exemple les informations nécessaires pour trouver le paquetage à déployer adéquat, le site cible, la forme binaire du paquetage. Ces produits sont typés(\textit{ProductType}) et circulent d'une activité à l'autre à travers des flots de données(\textit{Dataflow}).

Nous avons étendu ce modèle pour pouvoir intégrer les techniques de reprise sur erreurs décrites.Une activité est typée. Un point de reprise (\textit{Savepoint}) peut lui être associé. Il représente un état intermédiaire cohérent du procédé et contient les données nécessaires pour pouvoir retourner le procédé en arrière.

Nous introduisons également la notion de contexte d'activité. Le contexte d'une activité permet de récupérer  à tout moment des informations sur l'activité. Par exemple, il permet de récupérer au moment de la sortie en erreur des informations comme le taux du transfert pour une activité de transfert ou le pourcentage des fichiers copiés sur le disque dur pour une activité de déballage d'un paquetage. Selon la valeur de ces variables, on peut choisir une activité de reprise de l'erreur car pour certaines situations une stratégie de récupération peut être plus adéquate qu'une autre. Par exemple, pour l'activité de transfert, en fonction de la variable nommée, si sa valeur est de 80\% une contingence est plus adéquate qu'une compensation tandis que si sa valeur est de 10\% la compensation est plus indiquée.

\section{Validation}
Pour valider notre travail, nous avons expérimenté notre approche sur le système de déploiement ORYA\cite{ORYA3}. Il s'agit d'un système qui utilise une fédération d'outils progiciels existants pour offrir les fonctionnalités désirées et un moteur de procédés exécutables, APEL\cite{APEL}. Cet outil offre plusieurs fonctionnalités: il coordonne l'exécution des procédés, permet de définir et de modifier des procédés, permet de visualiser graphiquement l'exécution d'un procédé. 

\begin{figure*}[ht]
 \setlength{\fboxsep}{2mm}
 \fbox{
  \parbox{116mm}{
   \vspace{-20mm}
   \begin{center}
    \small
    \includegraphics[width=0.7\textwidth,angle=-90]{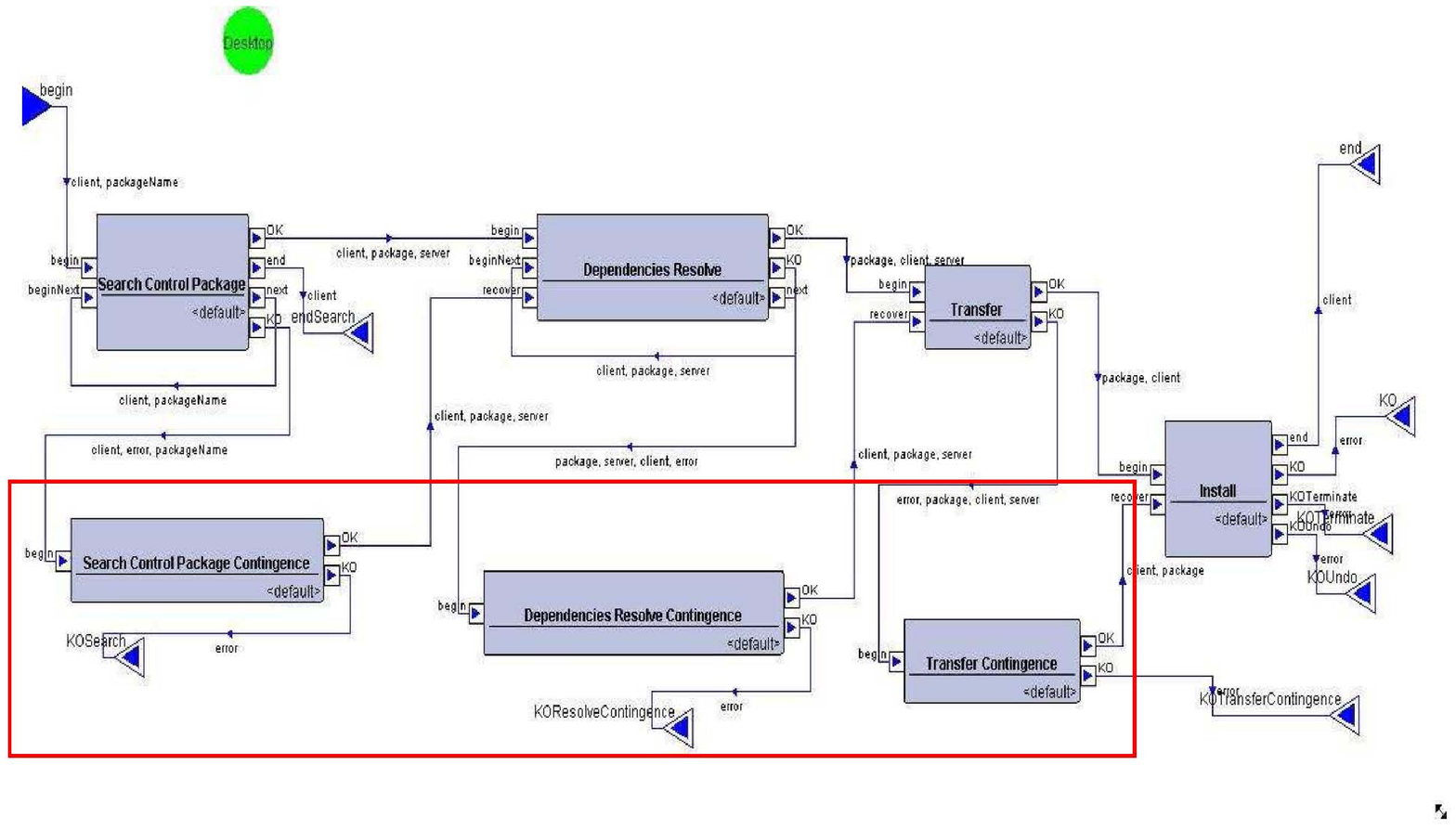}
   \end{center}
   \vspace{-20mm}
  }
 }
 \caption{Notre procédé d'installation}\label{fig:2}
\end{figure*}

ORYA offre un procédé de déploiement générique, qui peut être personalisé par ses utilisateurs. Le procédé d'installation offert par ORYA est constitué de plusieurs activités (la partie haute de l'image \ref{fig:2}). Pour pouvoir installer une application, le paquetage le contenant doit être trouvé (Search Control Package). Puis les dépendances logicielles de l'application sont résolues (Dependencies Resolve). Le paquetage est ensuite transféré (Transfert) vers le site ou il est installé (Install).

Pour valider notre approche, nous avons étendu le procédé générique que ORYA offre pour l'activité d'installation. L'extension consiste à récupérer les éventuelles erreurs/exceptions qui se produisent à l'exécution. Pour les résoudre, il y a deux politiques de traitement possibles. La première consiste à définir statiquement avant l'exécution les réactions qui doivent être exécutées. Notre définition statique du procédé d'installation transactionnel contient des activités de contingence connectées sur les ports de sortie en erreur des activités (port KO) (la partie base de l'image). Ainsi, chaque fois qu'une exception est levée, elle est envoyée vers le port KO de l'activité en cours d'exécution, d'où elle est transportée vers l'activité suivante. Celle-ci analyse l'erreur et en fonction du contexte exécute une contingence ou une compensation. 

La seconde se base sur une fonctionnalité du moteur de procédés APEL qui permet d'ajouter dynamiquement à l'exécution des activités. L'approche consiste à analyser les erreurs apparues et en fonction du contexte d'ajouter une activité de reprise (contingence ou compensation) pour résoudre le problème. Cette approche apporte un gain de complexité dans le système.

Nous sommes en cours d'expérimentation de notre procédé statique d'installation que nous avons présenté pour une application OSGi distribuée sur un parc de passerelles. Pour contrôler l'installation des \textit{bundles} de l'application sur chaque passerelle, un déployeur local est utilisé pour récupérer les éventuelles exceptions apparues et pour les transmettre vers le gestionnaire d'exécution. Le scénario d'exécution consiste en simuler une panne réseau entre le serveur de déploiement (ORYA) et une des passerelles, pour pouvoir appliquer une contingence qui consiste en l'utilisation d'une passerelle de réserve pour installer le \textit{bundle} correspondant.

\section{Conclusion}
Cet article présente une approche à base de concepts transactionnels pour résoudre les situations exceptionnelles apparues en cours d'exécution d'un procédé de déploiement. Ces concepts ont été introduits au niveau d'une activité qui compose un procédé de déploiement. 

Notre approche aide le système de déploiement à gagner en performance. Si une erreur apparaît, pour garantir la réussite du déploiement une activité de contingence peut être exécutée. Si cela n'est pas possible, la stratégie adoptée est de ne pas défaire tout le travail déjà effectué. Ainsi, le système de déploiement fait une retour partial en arrière pour assurer la sûreté du déploiement et pour économiser la ressource temps d'exécution.

Une des perspectives de ce travail que nous avons expérimenté que pour l'activité d'installation, est de compléter son validation pour les autres activités du cycle de vie ainsi que pour des situations nécessitant des protocoles plus difficiles de coordination (WS Coordinator,2PC). Une autre perspective de ce travail est de le personnaliser pour les différents modèles de plate formes dans le cadre du projet OSMOSE\cite{GDF}. 

\bibliography{DECOR04}

\end{document}